\documentclass[eqsecnum,nofootinbib,floats,preprint,aps,prd,floatfix,titlepage,
superscriptaddress]{revtex4} 

\usepackage{epsfig}
\usepackage{bm}
\usepackage{amsmath}

\def\high{\vphantom{\Biggl(}\displaystyle}

\begin{document}

\preprint{CU-TP-1185}
\preprint{KIAS-P08065}

\title{ BPS Magnetic Monopole Bags}
\author{Ki-Myeong Lee}
\email{klee@kias.re.kr}
\affiliation{School of Physics, Korea Institute for Advanced Study,
207-43, Cheongnyangni-2dong, Dongdaemun-gu, Seoul 130-722,
Korea}
\author{Erick J. Weinberg}
\email{ejw@phys.columbia.edu}
\affiliation{School of Physics, Korea Institute for Advanced Study,
207-43, Cheongnyangni-2dong, Dongdaemun-gu, Seoul 130-722,
Korea}
\affiliation{Physics Department, Columbia University, New York, New
York 10027, USA}

\vskip 1cm

\begin{abstract}
We explore the characteristics of spherical bags made of large numbers
of BPS magnetic monopoles.  There are two extreme limits.  In the 
Abelian bag, $N$ zeros of the Higgs field are arranged in a quasiregular 
lattice on a sphere of radius $R_{\rm cr} \sim N/v$, where $v$ is the Higgs vacuum
expectation value.  The massive gauge fields of the theory are largely
confined to a thin shell at this radius that separates an interior with 
almost vanishing magnetic and Higgs fields from an exterior region with 
long-range Coulomb magnetic and Higgs fields.  In the other limiting case,
which we term a non-Abelian bag, the $N$ zeros of the Higgs field are all the 
origin, but there is again a thin shell of radius $R_{\rm cr}$.  In this case
the region enclosed by this shell can be viewed as a large monopole core,
with small Higgs field but nontrivial massive and massless gauge fields.
\end{abstract}

\maketitle

\section{Introduction} 
\label{introduction}

There has recently been some interest in the nature of solutions
containing a large number of BPS monopoles.  Although we have a good
understanding of the field configurations when the number of magnetic
monopoles is relatively small, the situation becomes more complicated
when the number increases.  If the monopoles remain well separated,
the solution is, with an appropriate gauge choice, given approximately
by a superposition of individual BPS monopoles.  However, as the
number of monopoles increases, their non-Abelian cores increase in
size until they begin to overlap and the individual monopoles merge
together.

Bolognesi~\cite{Bolognesi:2005rk} has proposed that when $N \gg 1$ BPS
magnetic monopoles are as closely packed as possible in a single
region, the field configuration can be characterized as a bag, of
arbitrary shape, inside of which the Higgs field vanishes.  Following
this work, Ward~\cite{Ward:2006wt} explored numerically a ``monopole
wall'' on which BPS monopoles were arranged periodically on a plane.
The scalar field is approximately constant on one side of the wall,
but grows linearly with the distance from the wall on the other side.
This planar wall configuration can be regarded as an approximation to
the field configuration near the wall of a monopole bag.

However, much remains to be explored about the structure of these
solutions.  As long as $N$ remains finite (as it must for a finite
size monopole bag), the Higgs field, although exponentially small in
the bag interior, only vanishes exactly at a finite number of points.
Similarly, the charged vector fields are only precisely zero along
special lines with nontrivial vorticity.  By focusing on the patterns
of these zeros, and using previous 
work~\cite{Hitchin:1995qw,Houghton:1995bs,Houghton:1995bp,%
Houghton:1997kg,Sutcliffe:1996qz} on solutions with Platonic
symmetries as a guide, we will obtain a more detailed picture of the
monopole bag and its wall.  As we will see, the characteristics of the
solutions can change quite dramatically depending on the location of
the zeros of the Higgs field.

One limiting case occurs when most of the zeros are located on the
surface of the bag, the configuration one naturally obtains by
bringing many monopoles together from spatial infinity.  Up to
exponentially small corrections, the fields are purely Abelian both
inside and outside the bag, with the non-Abelian behavior confined to
the wall region; we will call these Abelian bags.  For a spherical
bag, a simple calculation shows that the minimum bag radius is of
order $N/v$, where $v$ is the expectation value of the Higgs field.
 At this radius, the non-Abelian cores of the monopoles
overlap, and the Higgs field is approximately zero inside the bag.
This gives Bolognesi's bag.  However one can also consider the case
where the monopoles are arranged on a surface with larger radius.  For
the spherical case, it is easy to see that the scalar field is approximately
constant, but nonvanishing, inside the bag.  A simple generalization of 
the magnetic conductor picture of Bolognesi allows one to see how to 
deform this configuration while keeping the Higgs field constant in
the interior.

Going in the opposite direction, one can move the Higgs zeros further
in toward the center.  This does not reduce the size of the bag, but
instead changes its character.  In the limiting case, where the zeros
all coincide at a point, the Higgs field remains close to zero inside
the bag, but the gauge field configuration becomes truly non-Abelian.
In a sense, the bag can be thought of as an extended monopole core.
We call this a non-Abelian bag.  

A remarkable feature of these bags is
that the bag wall has a lattice structure that gives the solution a
polyhedral shape.  For the Abelian bag, the vertices of the lattice
are defined, in an obvious manner, by the positions of the Higgs
zeros.  This clearly cannot happen for the non-Abelian bag, since the
zeros are all at the center.  We will see that a regular structure
emerges nevertheless, and will argue that it is an approximately
hexagonal lattice.

In the next section we will review the essential features of the
theory and establish our conventions.  Then, in Sec.~\ref{shellsection}, we 
discuss the general features of spherical bags, and describe the 
topological constraints on the zeros of the field.  We also
review here the properties of the Platonic monopoles.   In Secs.~\ref{abelsection}
and \ref{nonabelsection} we give more detailed pictures of the spherical Abelian and
non-Abelian bags, respectively.  Section~\ref{conclusion} contains some concluding 
remarks.

\section{Background and conventions}
\label{conventions}

We consider an SU(2) gauge theory with gauge field $V^a_\mu$ and a
triplet scalar field $\phi^a$.  We choose scales so as to set the gauge
coupling to unity.  The Lagrangian is
\begin{equation}
 {\cal L}=  -\frac{1}{4} G^a_{\mu\nu}G^{a\mu\nu} 
     + \frac12 D_\mu\phi^a D^\mu\phi^a   \,, 
 \end{equation}
where $D_\mu\phi^a = \partial_\mu\phi^a +\epsilon^{abc}V^b_\mu
\phi^c$, and $G^a_{\mu\nu} = \partial_\mu V^a_\nu-\partial_\nu V^a_\mu
+\epsilon^{abc}V^b_\mu V^c_\nu$.  We are interested in BPS solutions,
which obey the Bogomolny equation
\begin{equation}
   G^a_{ij}=\epsilon_{ijk}D_k\phi^a  \, .
\label{bogomolny}
\end{equation}

We will find it convenient for the most part to work in an
Abelian gauge where the Higgs field has a fixed gauge orientation,
\begin{equation}
   \phi^a = \delta^{a3} \phi  \, .
\end{equation}
The gauge field of the unbroken electromagnetic U(1) is then 
\begin{equation}
   A_\mu =  V_\mu^3   \, ,
\end{equation}
while the charged massive vector meson field is
\begin{equation}
   W_\mu = \frac{1}{\sqrt{2}}( V_\mu^1 + i V_\mu^2)   \, . 
\end{equation}
The components of the field strength are 
\begin{align}
 G^1_{\mu\nu}+iG^2_{\mu\nu} &= \sqrt{2}( {\cal D}_\mu W_\nu- {\cal D}_\nu W_\mu ) \cr
 G^3_{\mu\nu}   &= F _{\mu\nu} + i (W_\mu {W}^*_\nu-W_\nu{W}^*_\mu)  \, , 
\end{align}
where $F_{\mu\nu}=\partial_\mu A_\nu-\partial_\nu A_\mu$, and the electromagnetic 
covariant derivative ${\cal D}_\mu = \partial_\mu + iA_\mu $.

We will consider only static configurations with no electric charge,
and so assume that $A_0$ and $W_0$ both vanish everywhere.  We adopt three-vector
notation $\bf A$ and $\bf W$ for the spatial components of these fields and write
${\bf B} = {\bf \nabla \times A}$.
Equation~(\ref{bogomolny}) then becomes
\begin{align} 
  0 &=  {\bf B}+ i{\bf W}\times {\bf W}^* -\nabla \phi  
\label{Bequation}\\
  0 &=  {\cal D}\times {\bf W} +i{\bf  W}\phi   \, .
\label{Wequation}
\end{align}
The energy density is
\begin{equation}
   {\cal E} = {1\over 2}\left({\bf B} + i {\bf W \times W}^* \right)^2
             + \left| {\bf {\cal D} \times W} \right|^2
             + {1\over 2} (\nabla \phi)^2  +   |{\bf W}|^2 \phi^2  \, ,
\end{equation}
which for solutions of the Bogomolny equations becomes
\begin{equation}
   {\cal E} = \left({\bf B} + i {\bf W \times W}^* \right)^2
            +  2 |{\bf W}|^2 \phi^2   \, .
\end{equation}

We write the vector fields in terms of spherical coordinates, with
${\bf A}\cdot d{\bf r} = A_r dr + A_\theta d\theta + A_\varphi
d\varphi$ or, equivalently,
\begin{equation} 
  {\bf A}= \hat{r}A_r + \frac{ \hat{\theta}}{r} A_\theta+
  \frac{\hat{\varphi}}{r\sin\theta} A_\varphi   \, . 
\end{equation}
In particular, the vector potential for a Dirac monopole with magnetic
charge $N$ can be chosen so that $A_r=A_\theta=0$ and
\begin{equation}
    A_\varphi = N(1 - \cos \theta)   \, .
\end{equation}
We also define
\begin{equation} 
  W_\pm = (W_\theta\pm \frac{i}{\sin\theta} W_\varphi)
  e^{\pm i \varphi}    \, .
\end{equation}
The covariant curl of $W$ is then
\begin{align}   
   {\cal D}\times {\bf W} &= \frac{\hat{r}}{r^2\sin\theta}
   (\partial_\theta W_\varphi + iA_\theta W_\varphi -\partial_\varphi
   W_\theta -iA_\varphi W_\theta ) \nonumber \\ & 
   +\frac{\hat{\theta}}{r\sin\theta} (\partial_\varphi W_r +iA_\varphi
   W_r -\partial_r W_\varphi -iA_r W_\varphi ) \nonumber \\  & +
   \frac{\hat{\varphi}}{r}(\partial_r W_\theta +iA_rW_\theta
   -\partial_\theta W_r -iA_\theta W_r )    \, .
\end{align} 
Its radial component is given in terms of $W_+$ and $W_-$ via
\begin{align}   
    {\bf \hat{r}}\cdot {\cal D}\times {\bf W}  =& \frac{e^{-i\varphi}}{2r^2}
   \left\{ -i[\partial_\theta + iA_\theta ]
   -\frac{1}{\sin\theta}[\partial_\varphi + i
   (A_\varphi-1+\cos\theta)] \right\} W_+ \cr
   & + \frac{e^{i\varphi} }{2r^2 } \left\{
   i[\partial_\theta + iA_\theta ] -\frac{1}{\sin\theta}[\partial_\varphi
  + i (A_\varphi+1-\cos\theta ) ]\right\} W_-    \, .
\label{radialcurl} 
\end{align} 
We also note that 
\begin{equation}
   {\bf \hat r}\cdot \left({\bf B} + i {\bf W \times W}^* \right)
    =  B_r - \frac1{2r^2} \, |W_+|^2 + \frac1{2r^2}\,|W_-|^2   \, .
\label{radialBandW}
\end{equation}

\section{Shells, zeros, and Platonic monopoles}
\label{shellsection}

Consider the case of $N \gg 1$ monopoles distributed relatively evenly
on a spherical shell of radius $R$.  The typical separation between
neighboring monopoles is $d_{\rm sep} \sim 2\pi R/\sqrt{N}$, which we
initially assume to be much larger than the monopole core radius.
Outside the monopole cores, the only nontrivial fields are the Abelian
electromagnetic field and the massless scalar field with magnitude
$\phi({\bf x})$.  As long as we stay sufficiently far from the
monopole cores and the sphere on which they lie, and provided that we
take a sufficiently ``coarse-grained'' view of the system, we can
treat this as a spherically symmetric purely Abelian configuration.
Thus, the magnetic field will be
\begin{equation}
    {\bf B} \approx \begin{cases} \high  N \, {{\bf \hat r}\over r^2}
             \, , \quad   & \text {$ r \gtrsim R+a $ ,}  \cr
                \high  0  & \text{$ r \lesssim R-a  $.}\end{cases} 
\end{equation}
where we expect $a$, which measures the effective thickness of the spherical
shell containing the monopole cores, to be $\sim v^{-1}$.  Outside the monopole
cores the scalar field obeys ${\bf \nabla} \phi = {\bf B}$, with $\phi(r=\infty) = v$.
Hence,
\begin{equation}
    \phi \approx \begin{cases} \high  v - {N\over r}
               \, , \quad & \text{$ r \gtrsim R+a  $,}  \cr
        \high v - {N\over R}\, , \quad & \text{$ r \lesssim R-a  $.}
   \end{cases} 
\end{equation}

We see that there is a critical radius, $R_{\rm cr} \sim N/v$, for
which $\phi$ vanishes in the interior of the sphere.\footnote{Note
that $d_{\rm sep}(R_{\rm cr}) \sim \sqrt{N}/v$.  The monopole core
radius is set by the local value of $\phi$, and so only becomes
comparable to $d_{\rm sep}(R)$ when $(R - R_{\rm cr})/R_{\rm cr}$ is
of order $1/\sqrt{N}$.}  Once the sphere has shrunk to $R_{\rm cr}$,
the monopole cores merge together to give a thin spherical wall, 
within which the fields are fully
non-Abelian, that separates an essentially empty interior from a
purely Abelian exterior.  This is the Abelian monopole bag.  Because
$\phi$ cannot become negative, the bag cannot be shrunk beyond this
point.  The bag can be deformed from a spherical shape, but if it
remains spherical its radius cannot be less than $R_{\rm cr}$.

For this case of a bag with critical radius, our estimate of the wall
thickness needs to be modified.  This thickness is determined locally
by the properties of the Higgs field near the wall, where $\phi$ is
far from its vacuum value.  The only dimensional scale relevant near
the wall is the separation between monopoles on the wall, $d_{\rm
sep}(R_{\rm cr})$, so we expect the wall thickness to be of comparable
size; i.e., $\sim \sqrt{N}\,v^{-1}$.

Although the bag radius cannot be less than $R_{\rm cr}$, this
does not mean that the Higgs zeros cannot be brought closer together.
In fact, one can envision bringing them all to the center, giving an
$N$-fold zero and implying that $\phi \sim r^N$ near the origin.  For
large $N$, this means that we have a region with very small, although
not quite vanishing, Higgs field.  This region thus resembles an
enlarged monopole core, with both the massless gauge field and the
(just barely) massive gauge field being nontrivial.  How large is this
region?  In the exterior, at large $r$, the analysis above tells us
that $\phi \approx v - (N/r)$.  This behavior cannot continue for $r <
N /v$, so we again expect to find a spherical shell of radius $R_{\rm
cr} \sim N/v$ and thickness $\sim \sqrt{N}\,v^{-1}$ separating an exterior Abelian
region from an interior region of almost vanishing Higgs field.
However, in contrast with our previous case, this interior region is
not empty, but instead contains an intrinsically non-Abelian
configuration of fields.  This is the non-Abelian bag. 

In both cases, the fields in the exterior region can be expanded in
terms of spherical harmonics appropriate to magnetic charge $N$.  In
particular, the massive vector field can be expanded in terms of
monopole vector harmonics~\cite{Olsen:1990jm,Weinberg:1993sg}.  This
field falls exponentially for $R > R_{\rm cr}$, with the dominant part
of the exponential tail coming from the harmonics with the lowest
total angular momentum, $J=N-1$.  One can show~\cite{Ridgway:1995ke}
that any linear combination of such harmonics vanishes at precisely
$2N-2$ points on the unit sphere.  These correspond to directions in
which the multimonopole core region falls off faster, thus somewhat
``flattening'' the shape of the wall separating the exterior and
interior regions and deforming it from a sphere to a polyhedron with
$F=2N-2$ faces.

In fact, topological considerations imply the existence of other zeros
of the fields.  First, recall that the relation between the magnetic
charge and the winding number of the Higgs field implies that in an
$N$ monopole solution the number of zeros (with positive winding)
minus the number of antizeros (with negative winding) of the Higgs
field must be precisely $N$.  When the individual monopoles are widely
separated, it is clear that there are $N$ zeros and no antizeros.
However, this is not necessarily so; we will see that there are solutions
with $N$ units of magnetic charge that have $N+1$ Higgs zeros, with
the excess compensated by an antizero at the origin.

In addition, the various components of $\bf W$ are also required to
have zeros and, in fact, lines of zeros.  If the magnetic flux through
a surface $S$ is $4\pi P$, then any scalar field with unit electric
charge must have $2P$ zeros on $S$, with zeros being counted with a
positive (negative) sign if the phase of the field increases
(decreases) by $2\pi$ in going around the zero in a clockwise
direction, as seen by the outward normal.  The same is true of the
component of $\bf W$ normal to the surface.  The result is modified
for the tangential components.  In particular, on a sphere centered
about the origin $W_\pm$ have a total of $2P \mp 2$ zeros; the
additional factor in these cases can be understood by noting the extra
terms in Eq.~(\ref{radialcurl}) for the covariant curl.  By
considering concentric spheres of arbitrary radius, we see that these
zeros (and possibly antizeros) must form lines that can only terminate
on the points where the magnetic charges are located; i.e., at the
zeros of the Higgs field.  Note that this is consistent with our
previous remarks about the zeros of the vector harmonics with lowest
angular momentum, because these only contribute to $W_+$.

A connection between magnetic charge and polyhedral shape was first
encountered in a curious set of solutions that have the symmetries of
the Platonic solids, but not the charges that one might naively expect
to be associated with these solids.  Although in each case the energy
profile shows a concentration of energy at the vertices of a regular
polyhedron, the number of vertices is never equal to the magnetic
charge.  Thus, there is a charge-three
tetrahedron~\cite{Hitchin:1995qw,Houghton:1995bs}, a charge-four
cube~\cite{Hitchin:1995qw,Houghton:1995bs}, a charge-five
octahedron~\cite{Houghton:1995bp}, a charge-seven
dodecahedron~\cite{Houghton:1995bp}, and finally an icosahedron with
$N=11$~\cite{Houghton:1997kg}; for each of these the number of faces
is equal to $2N-2$.

The locations of the zeros of the Higgs field are rather
curious~\cite{Houghton:1995uj,Sutcliffe:1996qz}.  For the tetrahedron, octahedron,
and, it is believed\footnote{Although the existence of the N=11
icosahedral monopole solution has been established, its properties
have not be as fully explored as those of the other Platonic
solutions.}, the icosahedron, (i.e., the solids with triangular faces)
there are zeros at each of the $N+1$ vertices, and an antizero at the
center.  It is to be stressed that these should not be viewed as
simply arrangements of $N+1$ monopoles about a central antimonopole.
Such an assembly would be expected to have roughly $N+2$ times the
mass $M_1$ of a single monopole, whereas these solutions saturate the
BPS bound and so have mass $M_N =N M_1$.  It should also be noted that
plots~\cite{Sutcliffe:1996qz} of the Higgs field of the tetrahedron
and octahedron show that $\phi$ remains small throughout the interior
of the polyhedron.  For later use, we record their numbers of
vertices $V$, edges $E$, and faces $F$:
\begin{align}
       V &= N+1 \, , \cr
       E &= 3N - 3 \, , \qquad \qquad {\rm triangular~faces}   \cr
       F &= 2N - 2 \, .
\label{triangNumbers}
\end{align}

The energy profiles of the other two solutions, the cube and the
dodecahedron, also show concentrations about the vertices of the
polyhedron.  However, neither has Higgs zeros at these vertices.
Instead, they have multiple zeros --- four-fold and seven-fold,
respectively --- at their centers.  Again, plots of the Higgs field for
the cube~\cite{Sutcliffe:1996qz} and the dodecahedron~\cite{Paul} show
that it remains quite small throughout the interior region.  For these
solutions, corresponding to the duals of the previous
examples,\footnote{The tetrahedron is self-dual; note that
Eqs.~(\ref{triangNumbers}) and (\ref{dualNumbers}) agree when $N=4$.}
\begin{align}
       V &= 4N - 8 \, , \cr
       E &= 6N - 12 \, , \qquad \qquad {\rm dual~polyhedra}   \cr
       F &= 2N - 2 \, .
\label{dualNumbers}
\end{align}

\section{Spherical Abelian bags}
\label{abelsection}

As outlined in Sec.~\ref{shellsection}, we expect to be able to obtain an Abelian
monopole bag with charge $N\gg 1$ by arranging unit monopoles on a
spherical surface of critical radius.  Although we cannot have a
solution that is precisely spherically symmetric, we can hope to
obtain approximate spherical symmetry by arranging the monopoles in as
regular a fashion as possible.  

We expect to have concentrations of energy at the locations of the
monopoles, so that the solutions resemble polyhedra with their
vertices at the zeros of the Higgs field.  If we allow for the
possibility of a small number of antizeros or zeros located elsewhere,
this gives us $V=N+k$ vertices.  The arguments given previously
suggest that the number of faces should be $2N-2$.   Euler's theorem
then tells us that the number of edges is $E=3N+k-4 = 3F/2 + k-1$.
The smallest possible value of $k$ is unity, in which case all of 
the faces are triangular and there is a single Higgs antizero, presumably 
at the center.  This reproduces the result of Eq.~(\ref{triangNumbers}),
and suggests that the corresponding three Platonic solutions can be
seen as the prototypes of the Abelian bag solutions.

Let us now examine these solutions more closely, focusing on the
zeros of the components of the $\bf W$ field.  We begin by recalling
that on a sphere enclosing $P$ units of magnetic charge the
components $W_r$, $W_+$, and $W_-$ of the massive gauge field must
have $2P$, $2P-2$, and $2P+2$ zeros, respectively.  This can be
phrased more compactly by defining the helicity $\lambda$ to be 0, 1,
and $-1$ for $W_r$, $W_+$, and $W_-$, respectively; the number of zeros
for a given component of $\bf W$ is then $2(P-\lambda)$.
By considering a series of concentric spheres, we see that these
merge into lines of zeros that can only end at zeros or
antizeros of the Higgs field.  If we assume that the central Higgs antizero 
is located at the origin, the zero lines for the component of $\bf W$ with helicity
$\lambda$ can then include:
\begin{itemize}
\item{a)} $n_F^\lambda = (2N-2) k_F^\lambda$ lines from the origin,
through a face, to infinity,
\item{b)} $n_E^\lambda = (3N-3) k_E^\lambda$ lines from the origin,
through an edge, to infinity,
\item{c)} $n_V^\lambda = (N+1)  k_V^\lambda$  lines from a vertex to infinity,
\item{d)} $n_{V'}^\lambda = (N+1) k_{V'}^\lambda$ lines from the
origin to a vertex.
\end{itemize}
For the three cases of regular polyhedra, it is clear that 
these lines of zeros must all be radial
and must be arranged symmetrically.  Equation~(\ref{triangNumbers})
then implies that the various $k^\lambda_a$ are integers and give the
numbers of lines passing through an individual face, edge, or vertex.
Although the polyhedra with other values of $N$ do not have exact
symmetry, we expect a similar result to hold when $N$ is sufficiently
large.  We will assume this to be the case.

Consider a sphere centered at the origin and with radius large
enough that it encloses all of the Higgs zeros.  Because there are $N$
units of flux flowing outward through this sphere, $W_r$, $W_+$, and
$W_-$ must have $2N$, $(2N-2)$, and $(2N+2)$ zeros on the sphere,
respectively.  This implies that
\begin{equation}
   (2N-2) k_F^\lambda + (3N-3) k_E^\lambda + (N+1) k_V^\lambda
       = 2N - 2\lambda   \, .
\label{largesphereK}
\end{equation}
If we assume that the various $k_i^\lambda$ are independent of $N$,
this yields two conditions:
\begin{equation}
       2 k_F^\lambda + 3 k_E^\lambda +  k_V^\lambda =2 \, ,
\end{equation}
from the coefficients of $N$,
and
\begin{equation}
    -2 k_F^\lambda  -3  k_E^\lambda +  k_V^\lambda 
       = -2 \lambda  \, ,
\end{equation}
from the $N$-independent terms.
These imply that 
\begin{equation}
    k_V^\lambda  = 1 - \lambda  
\label{kVabelian}
\end{equation}
and
\begin{equation}
   2 k_F^\lambda + 3 k_E^\lambda = 1 + \lambda \, .
\label{faceEdge}
\end{equation}

Now consider a sphere centered at the origin, but lying inside the
zeros of the Higgs field.  Because this sphere encloses only the
antizero at the origin, it should have $-2$, $-4$, and 0 zeros of
$W_r$, $W_+$, and $W_-$, respectively; i.e., each component of $W$
must have $2N+2$ fewer zeros on this inner sphere than on the outer
one.  It immediately follows that $k^\lambda_{V'} = k^\lambda_V -2$,
so that
\begin{equation}
    k_{V'}^\lambda  = -1 - \lambda  \, . 
\label{kVprimeabelian}
\end{equation}

As a consistency check, consider a sphere that encloses a single
vertex, with one unit of flux flowing outward; we make take this
sphere to be small enough that only lines connected to a vertex run
through it.  To count zeros on this sphere, $W$ must be decomposed
into components $\tilde W_r$, $\tilde W_+$, and $\tilde W_-$ defined
with respect to its center; these must have 2, 0, and 4 zeros,
respectively.  On lines running radially outward to infinity from the
vertex, the $\tilde W_\lambda$ are the same as the $W_\lambda$ defined
with respect to the origin.  On the other hand, on lines running
radially into the center from the vertex the roles of $W_+$ and $W_-$
are interchanged.  Furthermore, the reversal in direction of the
outward normal changes the helicity, interchanging zeros and
antizeros.  The net result is the requirement that
\begin{align}
   k_V^0 - k_{V'}^0  &= 2 \, , \cr
   k_V^1 - k_{V'}^{-1}  &= 0 \, , \cr
   k_V^{-1} - k_{V'}^1  &= 4  \, .
\end{align}
These conditions are satisfied by the $k_V^\lambda$ and $k_{V'}^\lambda$
given by Eqs.~(\ref{kVabelian}) and (\ref{kVprimeabelian}).

The above arguments uniquely determine the $k_V^\lambda$ and
$k_{V'}^\lambda$, but leave some ambiguity as to the $k_F^\lambda$ and
$k_E^\lambda$.  This ambiguity can be resolved by choosing the
solutions of Eq.~(\ref{faceEdge}) that require the smallest total
number of zero lines; this criterion seems energetically reasonable,
since it would tend to minimize the angular derivatives of the fields.
With this choice, the numbers of zero lines of the various types are
given in Table~\ref{triangular}.

\begin{table}
\caption{Numbers of zero lines for solutions corresponding to polyhedra with 
triangular faces}
\begin{ruledtabular}
\begin{tabular}{ccccc}
 & {$k_F$} & $k_E$ & $k_V$ & $k_{V'}$  \\
 $W_+$ &  1 & 0 & 0 & -2   \\
 $W_r$ & -1 & 1 & 1 & -1   \\
 $W_-$ &  0 & 0 & 2 & 0  \\
\end{tabular}
\end{ruledtabular}
\label{triangular}
\end{table}  

Consider first the lines running through the vertices.
On the line running from the
vertex out to infinity, $W_-$ has a double zero and $W_r$ has a single
zero, while on the line connecting the origin and the vertex, $W_+$ has a
double antizero and $W_r$ again has a single antizero.  We can
understand these results in terms of energetic arguments.  The Higgs
zero at the vertex is a source for the magnetic field $\bf B$.  From
the symmetry of the configuration, we expect the outward magnetic
field to have a local maximum, as a function of angle, along the line
extending radially outward from the vertex.  Recalling
Eq.~(\ref{radialBandW}), we see that local energy density is reduced by
minimizing $|W_-|$ and making $|W_+|^2$ as close as possible to $B_r$.
$W_r$ does not enter explicitly here, but having a zero of $W_r$
allows a larger value of $|W_+|$ for a given value of $|{\bf W}|$, and
so is also energetically favored.  On the lines running from the
vertex toward the origin, the direction of $\bf B$ is inward.  This
interchanges the roles of $W_+$ and $W_-$.  It also reverses the
optimum helicity along the zero-lines, so that the zeros on the
external lines are replaced by antizeros on the internal ones.  

The behavior on the faces and along the edges also makes sense
energetically.  There is a zero of $W_+$ and an antizero of $W_r$ at
the center of each face, and a zero of $W_r$ at the midpoint of every
edge.  Since these are the points furthest from the zeros of the Higgs
field, one would expect $\phi$ to have a maximum there, which makes it
energetically favorable for $|{\bf W}|$ to be small.  This is also
consistent with the arguments given in Sec.~\ref{triangular} that the
$J=N-1$ vector harmonics (which contribute only to $W_+$)
should have a zero on each face.

Our discussion has been carried out in a gauge where the Higgs field
has a fixed SU(2) orientation.  (This implies the existence of Dirac
strings, but these do not affect any of our arguments.)  However, one
can also work in a nonsingular gauge where the winding number of the
asymptotic Higgs field is given by the magnetic charge.  Although the
Higgs field orientation is a gauge-variant quantity and need not be
symmetric, let us consider what a symmetric (or almost symmetric)
Higgs field would look like.  If we were to ignore the antizero at the
center, all of the Higgs winding would have to occur outside the
sphere of radius $R_{\rm cr}$ on which the Higgs zeros lie.  Let us
therefore consider the Higgs field on a sphere in this exterior
region.  As $N$, and thus $R_{\rm cr}$, becomes large, a region of
fixed transverse size approaches a plane.  A symmetric Higgs winding
could then be obtained by taking $\hat \phi^a =\phi^a/|\phi|=
\delta^{a3}$ above each zero of the Higgs field (i.e., at each of the
vertices of the triangulation of the sphere).  The lattice dual to
this triangulation is a hexagonal one, with the edges of the hexagons
being the lines joining the centers of the triangles.  To make the
total winding number match the magnetic field, the Higgs field should
take the opposite orientation, $\hat \phi^a = -\delta^{a3}$ on these
hexagonal edges.  However, this prescription is exact only in the
$N\to \infty$ planar limit.  With finite $N$, and a spherical shape,
the Higgs orientation should vary slightly from vertex to vertex.
Instead of the above prescription, we should have $\hat \phi^a = \hat
r^a$ at each of the vertices, with $\hat \phi^a$ taking the opposite
orientation along the corresponding hexagon (and the occasional
pentagon) of the dual lattice.  This variation from vertex to vertex
reduces the Higgs winding slightly, so that the total winding number
is one less than the number of vertices.  This is the origin of the
Higgs antizero at the center.

\section{Spherical non-Abelian magnetic bags}
\label{nonabelsection}

We now turn to the non-Abelian bag, focusing on the extreme case
where all of the Higgs zeros coincide at the origin.  For the Abelian
bag, the presence of a finite number of Higgs zeros on the bag wall
clearly ruled out the possibility of exact spherical symmetry.  One
might have thought that putting the zeros all at the same point would
allow such symmetry for the non-Abelian case, were it not that it was
shown long ago~\cite{Weinberg:1976eq} that spherical symmetry is only
allowed for $N= 1$.  One way to understand this result is to observe
that the components of $\bf W$ each have finite numbers of zero lines,
whose locations necessarily break the spherical symmetry.\footnote{The
$N=1$ monopole solution can be spherically symmetric because $W_+$ has
no zero lines (since $2N-2=0$), while $W_r$ and $W_-$ vanish
everywhere.}

For the Abelian bag, the shape of the bag wall and the polyhedral
structure on that wall were given directly by the locations of the
Higgs zeros.  In the non-Abelian case, the features determining the
polyhedral structure are much less obvious.  To put this in terms of
collective coordinates, recall that the charge $N$ solution depends on
$4N-1$ parameters, which can be taken to be the positions and U(1)
phases of the component monopoles, less one overall global phase.  For
the Abelian bag, even after the positions of the Higgs zeros have been
chosen to define its gross structure, there are still roughly $N$
phase variables that can be adjusted.  For the non-Abelian bag, on the
other hand, requiring that the Higgs zeros all lie at the origin leaves only
$N-1$ adjustable parameters, which is far less than would be needed to
specify the vertices of an arbitrary polyhedral bag.

Of course, one might ask whether it is even possible to have an
arbitrary number of Higgs zeros coincide.  The cubic and dodecahedral
solutions show that this can be done for $N=4$ or 7, and axially
symmetric solutions with multiple zeros at the origin can be
constructed for arbitrary $N$.  Note, though, that the latter are
toroidal in shape, and for large $N$ are more disklike than
polyhedral.  Although we see no reason why polyhedral solutions should
not exist for arbitrary large $N$, we have no rigorous proof, but must 
assume that they do.

As with the Abelian bags, we expect the $2N-2$ zeros of the lowest,
$J=N-1$, harmonics of $\bf W$ to lead to a polyhedral shape with
$2N-2$ faces.  In contrast with the Abelian case, where the locations
of these faces, including their distances from the center, were
uniquely determined by the Higgs zeros, there are now only lines of
zeros emanating from the origin.  Assuming these lines to be radial,
we will have a foliation of space by concentric polyhedra, but with no
specific polyhedron uniquely picked out, although the ones lying
within the thin wall region in which $|{\bf W}|$ rapidly goes to zero are
clearly special.   

Recalling Eq.~(\ref{radialBandW}), we see that it is energetically
favorable for $\bf B$ to be largest in the directions where $W_+$ is
largest.  Since the latter vanishes at the center of each face, it
seems likely that the magnetic flux will be greatest along the
directions corresponding to the vertices.  To spread this flux out as
evenly as possible, one would want to maximize the number of vertices
for a given value of $N$.  This is done by requiring that each vertex
be trivalent; i.e., that precisely three edges emanate from each
vertex.  Hence, we expect the non-Abelian polyhedra to be the duals of
the triangular polyhedra of the Abelian bags; this is consistent with
the cubic and dodecahedral examples.  The numbers of vertices, edges,
and faces are then given by Eq.~(\ref{dualNumbers}).  For large $N$,
the most symmetric solution would then have surfaces that were mostly
covered by approximately regular hexagons, with exactly twelve
pentagons distributed among them.\footnote{The topological constraints
can also be satisfied by replacing pairs of pentagons by
quadrilaterals, or triplets of pentagons by triangles, as in
the cube and the tetrahedron.}

We now turn to the lines of zeros associated with the components of $\bf W$,
proceeding as we did for the Abelian case.  There will be 
\begin{itemize}
\item{a)} $n_F^\lambda = (2N-2) \ell_F^\lambda$ lines from the origin,
through a face, to infinity ,
\item{b)} $n_E^\lambda = (6N-12) \ell_E^\lambda$ lines from the origin,
through an edge, to infinity,
\item{c)} $n_V^\lambda = (4N-8) \ell_V^\lambda$ lines from a vertex to
infinity,
\end{itemize}
with the $\ell_a^\lambda$ all integers.   

By considering any sphere centered about the origin we obtain the analog of 
Eq.~(\ref{largesphereK}),
\begin{equation}
   (2N-2) \ell_F^\lambda + (6N-12) \ell_E^\lambda + (4N-8) \ell_V^\lambda
       =  2N - 2\lambda   \, .
\label{dualsphereL}
\end{equation}
Proceeding as before, we assume that, at least for large $N$,
the $\ell_a^\lambda$ are
independent of $N$.  Equation~(\ref{dualsphereL}) then gives two
equations, which are solved by 
\begin{equation}
       \ell_F^\lambda =  2 - \lambda
\end{equation}
and 
\begin{equation}
      3 \ell_E + 2 \ell_V =  -1 + \lambda \, .
\label{dualambig}
\end{equation}
If we again take the solutions that require the fewest zero lines, we
obtain the results shown in Table~\ref{dualtable}.

\begin{table}
\caption{Numbers of zero lines for solutions corresponding to polyhedra with
hexagonal and pentagonal faces}
\begin{ruledtabular}
\begin{tabular}{cccc}
 & {$\ell_F$} & $\ell_E$ & $\ell_V$   \\
 $W_+$ &  1 & 0 & 0    \\
 $W_r$ & 2 & -1 & 1    \\
 $W_-$ &  3 & 0 & -1   \\
\end{tabular}
\end{ruledtabular}
\label{dualtable}
\end{table}

The results for the lines of zeros through the vertices are similar to those
for the Abelian case, and are consistent with the magnitude of $\bf B$
having a maximum in these directions.  The zero of $W_-$
decreases the right-hand side of Eq.~(\ref{radialBandW}), while the
zero of $W_r$ allows $|W_+|^2$ to be larger, again decreasing this
contribution to the energy.  The multiple zeros of $W_r$ and $W_-$ in 
the faces strongly suggest that $\phi$ has minima at the centers of the 
faces.

\section{Concluding remarks}
\label{conclusion}

The examples we have studied are not the only possible quasispherical
solutions.  One could certainly bring together a collection of monopoles
to create an Abelian bag with a hexagonal, rather than
triangular, lattice.  However, this lattice structure would be less
regular, with zeros of $W_+$ on only some faces.  Alternatively, one
could construct a lattice with more than one monopole at each vertex.

There will also be hybrid bags, with some Higgs zeros in the bag wall
and some in the interior.  Consider for example, a solution with $N/2$
zeros arranged symmetrically around a sphere of radius $R_2=N$ and
$N/2$ zeros arranged around a sphere of radius $R_1 < R_2$.  For $r<
R_1$ this would resemble the Abelian bag, but for $R_1 < r < R_2$ the
fields would be similar to those of the non-Abelian bag.

It is unclear, however, what the range of possibilities is for the non-Abelian
bag, especially for the case where the Higgs zeros are all at the
origin.  Here, the lattice structure emerges from the nonlinear
interactions of the fields, rather than being imposed by the location
of Higgs zeros at the surface of the bag.  As we have already noted, the 
number of parameters that can be varied is far fewer than would be needed
to specify an arbitrary lattice on the bag wall.

There can also be nonspherical bags.  One with an Abelian interior can
be constructed by assembling an irregular array of separated monopoles
and bringing them together just till the point where their cores
overlap and $\phi$ becomes exponentially small in the interior.  This
requires a nonuniform density of zeros on the bag wall, with the
density being highest in the regions where the wall curvature is the
greatest.  More precisely, the distribution of zeros can be obtained
by using the magnetic conductor picture of Bolognesi, and requiring
that the monopole density yield a constant dual magnetic potential in
the interior of the bag.  In fact, it is clear from this approach that,
by reducing the monopole density on the bag wall, one can also obtain a
nonspherical bag with a nonzero, but approximately uniform, Higgs
field in its interior.

Finally, note that as $N$ and, consequently, the bag radius become
large, any finite region of the bag wall can be approximated by a
plane, which we may take to be the $x$-$y$ plane.  In the case of the
Abelian bag, the wall would be similar to the monopole wall studied by
Ward~\cite{Ward:2006wt}, except that it would contain a triangular,
rather than a square, lattice of Higgs zeros.  On the side of the wall
corresponding to the bag interior (which we choose to be $z<0$), the
magnetic field and the Higgs field would rapidly approach zero with
increasing distance from the wall.  On the other side, the variation
of $\bf B$ and $\phi$ with $x$ and $y$ would tend to vanish with
distance from the wall.  The former would approach a constant vector
$B_i= \delta_{i3} B$, while asymptotically the latter would increase
linearly, with $\phi \approx B z$.

Although $\bf W$ would be exponentially small outside the wall, it
would not be precisely zero, and so the lines of zeros would persist.
The analog of $W_r$ would be $W_z$, while $W_x \pm i W_y$ would
correspond to $W_\pm$.  In the planar limit
the constraints on the zeros of all three of these components are all the
same.  For a triangular lattice with $E= 3F/2 = 3V$, the analog of
Eq.~(\ref{largesphereK}) in the region $0<z<\infty$, where there is a
nonzero magnetic flux, is
\begin{equation}
      2k_F + 3k_E + k_V = 2  \, .
\end{equation} 
There is no net magnetic flux through any plane with negative $z$,
so for $-\infty < z < 0$ we have
\begin{equation}
      2k_F + 3k_E + k_{V'} = 0 \, ,
\end{equation}
where $k_{V'}$ refers to lines running from $-\infty$ to a vertex.
Because lines of zeros can only end at a zero of the Higgs field,
$k_F$ and $k_E$ must be continuous in going though the wall.  These
equations are satisfied by the values in Table~\ref{triangular} as, of
course, they must.

Several possible variations on this picture immediately come to mind.
One possibility is to have $\phi$ asymptotically constant, but
nonzero, as $z \rightarrow \infty$; this would be the planar limit of
an Abelian bag with radius greater than $R_{\rm cr}$.  Corresponding
to the case of nested shells of zeros would be a series of parallel
planes, each with its own lattice of Higgs zeros and monopole cores.
For these solutions, $\phi$ is approximately linear in $z$, but with a
slope that changes as one passes through each wall.  Finally, there
are ``monopole sheet'' solutions~\cite{Lee:1998isa,Ward:2005nn} in
which $B$ points outward on both sides of the wall; these cannot be
realized as limits of a bag solution.

The case of the non-Abelian bag is perhaps more challenging, because
there are no Higgs zeros to fix the location of the wall.
Nevertheless, we can understand how a wall can arise by making some
simplifying assumptions.  To start, let us consider fields that have
been averaged over a lattice plaquette in the $x$ and $y$ directions,
so that they are independent (up to a gauge transformation) of these
variables, and only have nontrivial dependence on $z$.  It follows
that $\bf B$ is constant and has only a $z$ component, and hence that
we can set $A_z =0$.  Let us assume that $W_z$ also vanishes.
Finally, let us assume that the behavior of $W_x$ and $W_y$ is similar
to that of the lowest monopole harmonics, so that $W_x = i W_y \equiv
w(z)$.

With these assumptions, Eqs.~(\ref{Bequation}) and (\ref{Wequation})
yield two nontrivial equations,
\begin{equation}
     {d\phi \over dz} = B - 2|w|^2 
\end{equation}
and 
\begin{equation}
     {dw \over dz} = -\phi w  \, .
\label{Wprime}
\end{equation}
Differentiating the first of these and then using the original two equations to
eliminate $w$ leads to 
\begin{equation}
  0=  {d^2\phi \over dz^2}  + 2 \phi  {d\phi \over dz} - 2 B \phi \, .
\label{phiprimeprime}
\end{equation} 

\begin{figure}
\begin{center}
\leavevmode
\epsfysize=3.5in
\epsffile{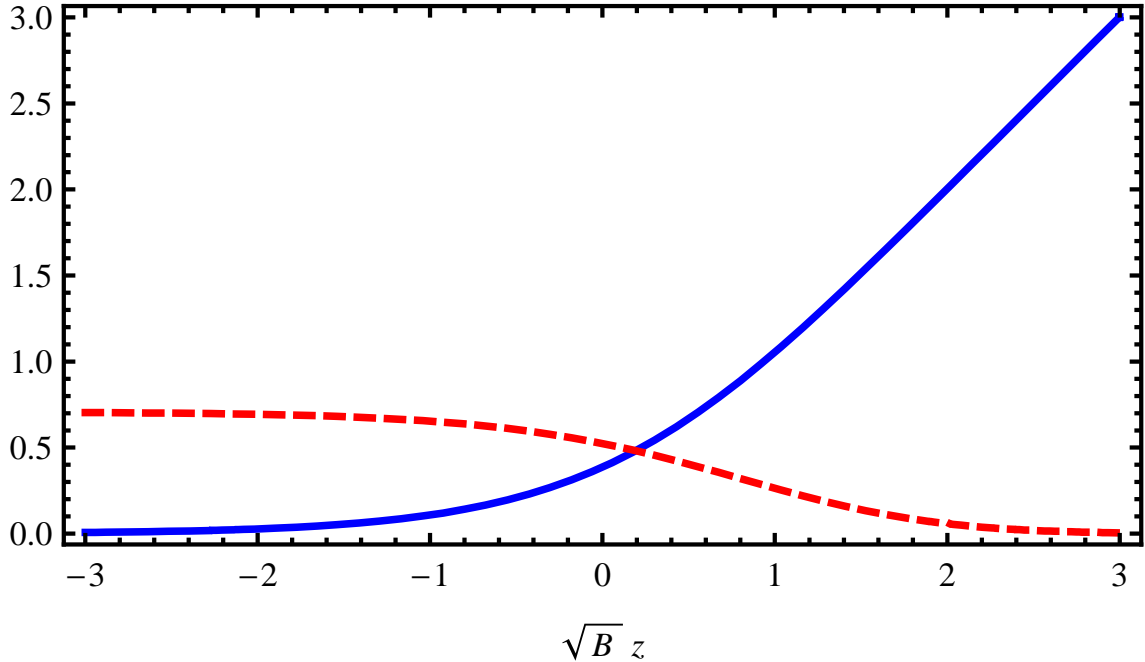}
\end{center}
\caption{The functions $w/\sqrt{B}$ (dashed red line) and
$\phi/\sqrt{B}$ (solid blue line)for the planar solution
corresponding to the large $N$ limit of the non-Abelian bag.}
\label{nonabelfig}
\end{figure}

In a generic solution of this equation, $|\phi|$ diverges at both
$-\infty$ and $\infty$.  However, there are also solutions with the 
asymptotic behavior
\begin{equation} 
     \phi \approx \begin{cases} B (z -z_0) \, , & 
        \quad  \text{$ z \rightarrow \infty$} , \cr
          e^{\sqrt{2B} \,  z}  \, , 
        & \quad  \text{$ z \rightarrow -\infty$}.  
       \end{cases}
\end{equation} 
that match our expectations for the wall of the non-Abelian bag; the
dependence on the arbitrary parameter $z_0$ reflects the translation
invariance of the theory.  Once $\phi(z)$ is known, Eq.~(\ref{Wprime}),
together with the fact that $d\phi/dz$ vanishes at $z = -\infty$,
gives
\begin{equation}
    w(z) = 
   \sqrt{B}\, \exp\left[-\int_{-\infty}^z dz' \, \phi(z') \right] \, .
\end{equation}
These solutions are shown in Fig.~\ref{nonabelfig}.  The wall region, in which
$w$ transitions from being essentially constant to being exponentially
small, is clearly evident.  A rescaling of variables in
Eq.~(\ref{phiprimeprime}) shows that this has a width of order
$B^{-1/2}$.  This agrees with our estimate, in
Sec.~\ref{shellsection}, that the wall of a spherical bag with
critical radius should have a thickness $\sim \sqrt{N} v^{-1} \sim
[B(R_{\rm cr})]^{-1/2}$.

To sum up, in this paper we have explored BPS monopole solutions in
the limit of large magnetic charge.  Our results have confirmed and
refined the bag picture and elucidated its fine structure, and have
shown that the monopole bag can be realized in a novel non-Abelian
manner.

\begin{acknowledgments} 

We thank Paul Sutcliffe for very helpful and informative discussions.
This work was supported in part by the Korea Research Foundation,
KRF-2006--C00008, by the KOSEF SRC Program through CQUeST at Sogang
University, and by the U.S. Department of Energy.

\end{acknowledgments}

\end{document}